\documentclass{iaus}
\usepackage{graphicx}

\title[Eclipsing binaries in the SMC] 
{New distance and depth estimates from observations of eclipsing binaries in the SMC}

\author[Pierre L. North, Romain Gauderon \& Fr\'ed\'eric Royer]   
{Pierre L. North$^1$, Romain Gauderon$^1$
 \and Fr\'ed\'eric Royer$^2$}

\affiliation{$^1$Ecole Polytechnique F\'ed\'erale de Lausanne (EPFL), \\ Observatoire de Sauverny,
CH-1290 Versoix, Switzerland \\ email: {\tt pierre.north@epfl.ch} \\[\affilskip]
$^2$GEPI, Observatoire de Paris -- Section de Meudon, \\ 5, place Jules Jansen,
F-92195 Meudon Cedex, France \\email: {\tt Frédéric.Royer@obspm.fr}}

\pubyear{2008}
\volume{256}  
\pagerange{119--126}
\jname{The Magellanic System: Stars, Gas, and Galaxies}
\editors{Jacco Th. van Loon \& Joana M. Oliveira, eds.}
\begin{document}

\maketitle

\begin{abstract}
A sample of 33 eclipsing binaries observed in a field of the SMC with FLAMES@VLT is presented.
The radial velocity curves obtained, together with existing OGLE light curves, allowed the
determination of all stellar and orbital parameters of these binary systems. The mean distance
modulus of the observed part of the SMC is 19.05, based on the 26 most reliable systems. Assuming
an average error of 0.1~mag on the distance modulus to an individual system, and a gaussian
distribution of the distance moduli, we obtain a $2-\sigma$ depth of 0.36~mag or 10.6~kpc.
Some results on the kinematics of the binary stars and of the H{\textsc ii} gas are also given.
\end{abstract}

\firstsection 
\section{Introduction}

The last decade has seen a renewal of interest for eclipsing binary stars,
thanks to the release of a huge number  of light curves as a byproduct of automated
microlensing surveys (EROS, MACHO, OGLE etc) with 1-m class telescopes . The reader
can refer to the reviews from Clausen (\cite{jC04}) and Guinan (\cite{eG04},
\cite{eG07}). The interest of eclipsing binary systems reside essentially on the
potential they offer to determine with excellent accuracy the masses and the radii
of the stellar components. To that end, both photometric and spectroscopic (radial
velocity) data are needed. If the metallicity is known, and if the surface brightness
of each component is well determined (through spectroscopic or photometric estimate
of effective temperature), tests of evolutionary models of single stars can be made,
provided the components are sufficiently far apart (``detached'' systems). Such tests
have been discussed by e.g. \cite[Andersen (1991)]{jA90}.

Conversely, one can consider the internal structure models as reliable enough, and use
them to determine both the metallicity $Z$ and helium content $Y$, which give access
to the relative enrichment $\Delta Y/\Delta Z$. That original approach was proposed
by \cite[Ribas et al. (2000a)]{RJ00}; their sample included essentially Galactic binary
systems, with only one belonging to the LMC. Of course, the slope
$\Delta Y/\Delta Z$ would be much better constrained by adding a large number of SMC
systems.

Another reason to focus on eclipsing binaries in the Magellanic Clouds is that we have a
nearly complete sample of such objects to a given limiting magnitude, making them
representative of a whole galaxy. Therefore, statistics of the orbital elements of
detached systems can potentially yield clues about the formation mechanisms of such
systems, and the study of semi-detached and contact ones may constrain scenarios of
binary evolution. Here, we will briefly discuss the ``twin hypothesis''
(\cite[Pinsonneault \& Stanek 2006]{PS06}) which suggests an excess of systems with a
mass ration close to 1.

Until a purely geometrical distance determination is feasible, 
Paczy\'{n}ski (\cite{bP01}) considered that detached EBs are the
most promising distance indicators to the Magellanic Clouds. 
Four B-type EB systems belonging to the Large Magellanic 
Cloud (LMC) were accurately characterized in a series of papers 
by Guinan et al. (\cite{GFD98}), Ribas et al. (\cite{RGF00}, \cite{RFMGU02}) and
Fitzpatrick et al. (\cite{FRG02}, \cite{FRGMC03}).
More recently, from high resolution, high $S/N$ spectra obtained with UVES at the ESO VLT,
the analysis of eight more LMC systems was presented by Gonz\'{a}lez et al. (\cite{GOMM05}). 
Harries et al. (\cite{HHH03}, hereinafter HHH03) and Hilditch et al.
(\cite{HHH05}, hereinafter HHH05) have given the 
fundamental parameters of a total of 50 EB systems of spectral types O and B. 
The spectroscopic data were obtained with the 2dF multi-object spectrograph on the 
3.9-m Anglo-Australian Telescope. This was the first use of 
multi-object spectroscopy in the field of extragalactic EBs. Let us also
mention that the distances of an EB in M31 (\cite[Ribas et al. 2005]{RJVFHG05}) and
another in M33 (\cite[Bonanos et al. 2006]{BSK06}) were measured recently.    
Although the controversy about the distance to the Magellanic Clouds seems to be solved in
favour of a mid position between the ``short"
and the ``long" scales, distance data and line of sight depth remain vital for comparison with
theoretical models concerning the three-dimensional structure and the kinematics of the SMC
(Stanimirovi\'{c} et al. \cite{SSJ04}).

Our contribution provides both qualitative and quantitative improvement over previous
studies. Thanks to the VLT GIRAFFE facility, spectra were obtained with a resolution
three times that in Hilditch's study. Another strong point is the treatment of nebular
emission. The SMC is known to be rich in H\,\textsc{ii} regions (Fitzpatrick \cite{elF85},
Torres \& Carranza \cite{TC87}). Thus, strong emission frequently appears superposed to
photospheric Balmer lines, which complicates the analysis.

\section{Observations}

\begin{figure}[b]
\begin{center}
\includegraphics[height=3cm]{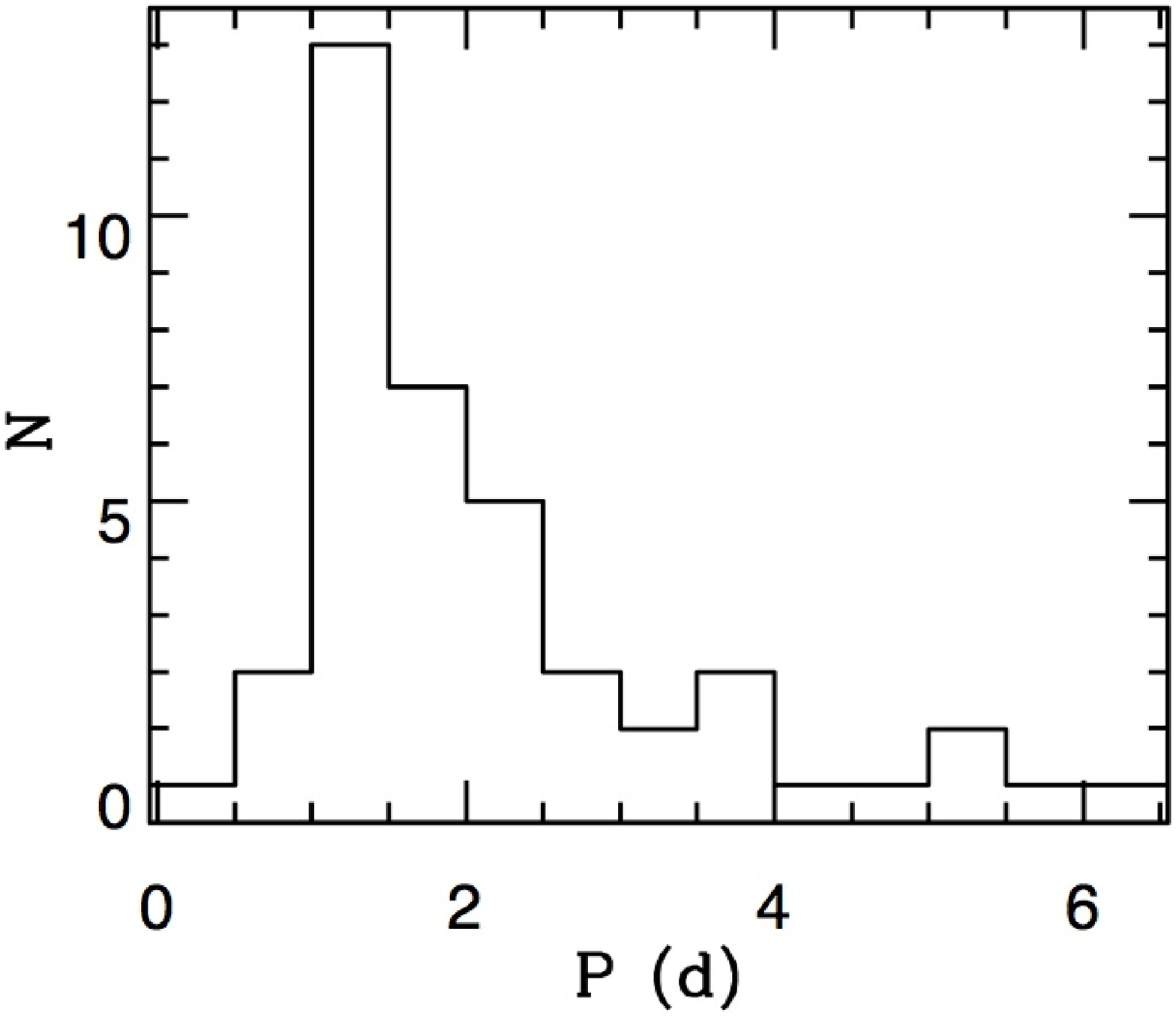}
\includegraphics[height=3cm]{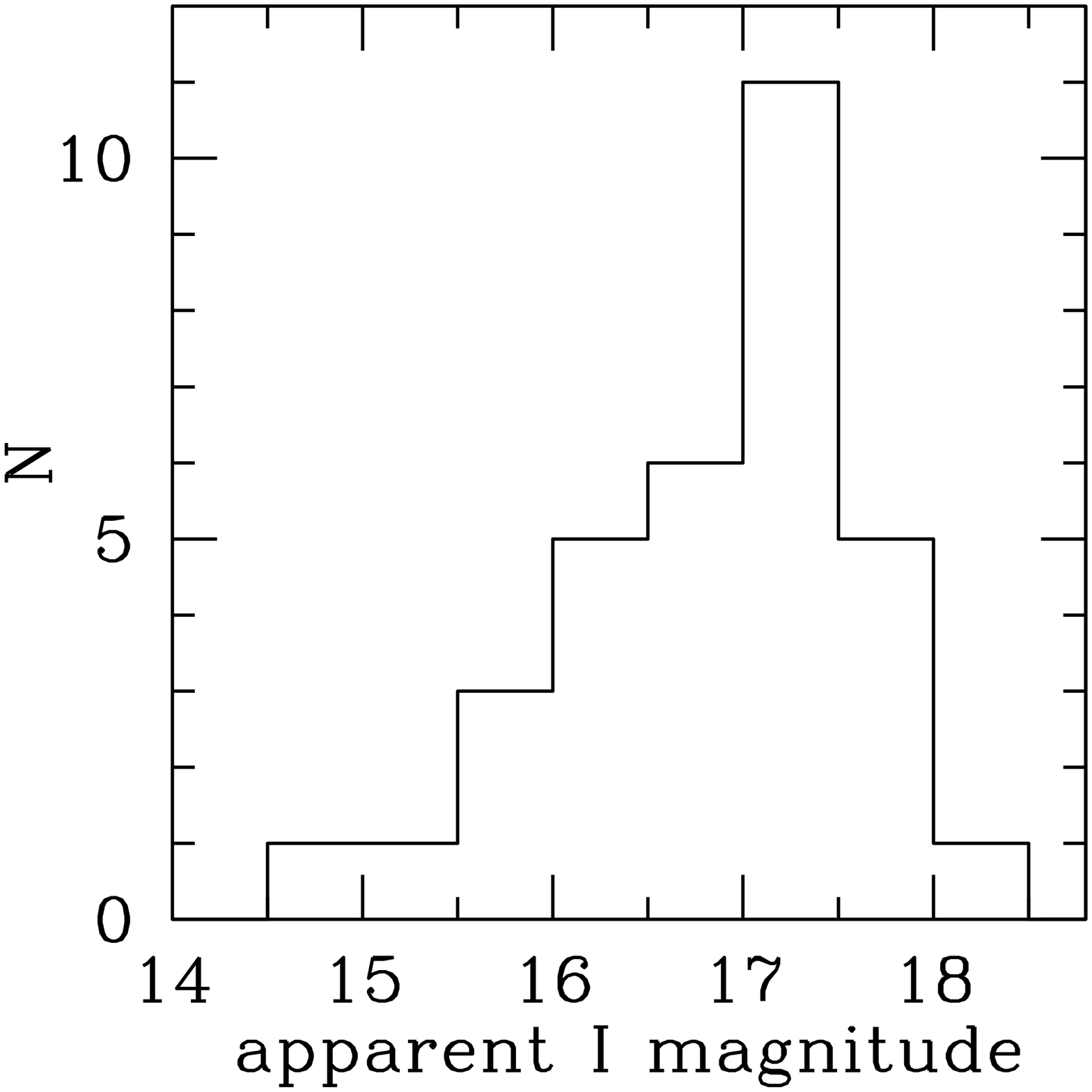}
\caption{Left: Histogram of periods of our sample of 33 eclipsing binaries in
0.5 day bins. Right: Histogram of apparent I magnitudes at quadrature.}
\label{periods}%
\end{center}
\end{figure}

The targets, astrometry included, were selected from the first OGLE photometric catalog. The
GIRAFFE field of view (FoV) constrained to choose systems inside a 25' diameter circle.
Other constraints were $I\leq18$ mag, at least 15 well-behaved detached light curves
and finally seven bump cepheids in the FoV (for another program). The
observations were done in November 2003 during eight consecutive nights, and the field
was observed twice a night, which makes 16 observations in all. The exposure time was 43 minutes
for all but one exposure which was limited to 12 minutes because of a technical problem.
Fig.\,\ref{periods} shows the histogram of the orbital periods of the 33 systems. 

For all but two binaries, the light curves come from the new version of the OGLE-II catalog of
eclipsing binaries detected in the SMC (Wyrzykowski et al. \cite{WUK04}). This catalog is based
on the Difference Image Analysis (DIA) catalog of variable stars in the SMC. The data were
collected from 1997 to 2000. Two systems were selected from the first version of the catalog
(standard psf photometry) but for an unknow reason they do not appear any more in the new
version.
 
The DIA photometry is based on $I$-band observations (between 202 and 312 points per curve).
$B$ and $V$ light curves were also used in spite of a much poorer sampling (22--28
points/curve and 28--46 points/curve in $B$ and $V$ respectively).
The objects studied in this paper have an average $I$ magnitude and scatter
(calculated from the best-fitting synthetic light curves) in the range
$15.083\pm0.009$ to $18.159\pm 0.047$.
 
The spectrograph was used in the low resolution (LR2) Medusa 
mode: resolving power $R=6400$, bandwidth $\Delta\lambda=603$~\AA\ centered on
$4272$~\AA. 
The most prominent absorption lines in the blue part of early-B stars spectra are:
H$\epsilon$, He\,\textsc{i} $\lambda4026$, H$\delta$, He\,\textsc{i} $\lambda4144$, H$\gamma$,
He\,\textsc{i} $\lambda4388$, and He\,\textsc{i} $\lambda4471$. For late-O
stars, He\,\textsc{ii} $\lambda4200$ and He\,\textsc{ii} $\lambda4542$ gain in importance. 

Beside the spectra of the objects, 21 sky spectra were obtained for each exposure in
the SMC.

\section{Data reduction and analysis}

The basic reduction and calibration steps including velocity correction to the heliocentric 
reference frame for the spectra were performed with the GIRAFFE Base Line Data 
Reduction Software (BLDRS). We subtracted a continuum component of the sky, obtained from 
an average of the 21 sky spectra measured over the whole FoV. For a given epoch the sky level
varies slightly across the field, but we neglected that second order correction.
Normalization to the continuum, cosmic-rays removal and Gaussian smoothing ($FWHM$ = 3.3 pix)
were performed with standard NOAO/PyRAF tasks.   
Spectral disentangling was performed with the KOREL code (Hadrava \cite{pH95}, \cite{pH04}),
which also gives the radial velocities and orbital elements. 
The simultaneous analysis of light curves and RV curves was made with the 2003 version of the
Wilson-Devinney (WD) Binary Star Observables Program (Wilson \& Devinney \cite{WD71};
Wilson \cite{rW79}, \cite{rW90}) via the PHOEBE  interface (Pr\v{s}a \& Zwitter \cite{PZ05}).

{\underline{\it Radial velocities}}.
Simon \& Sturm (\cite{SS94}) were the first to propose a method allowing the simultaneous
recovery of the individual spectra of the components and of the radial velocities. Another
method aimed at the same results, but using Fourier transforms to save computing time,
was proposed almost simulatneously by Hadrava (\cite{pH95}). The advantages of these
methods are that they need no hypothesis about the nature of the components of the binary
system, except that their individual spectra
remain constant with time. Contrary to the correlation techniques, no template is needed.
In addition to getting at once the radial velocities and orbital elements, one gets the
individual spectra of the components (``disentangling''), with a signal-to-noise ratio
which significantly exceeds that of the observed composite spectra.
Other details about these techniques and their
applications can be found in e.g. Hensberge et al.
(\cite{HPV00}), Pavlovski \& Hensberge (\cite{PH05}) and Hensberge \& Pavlovski
(\cite{HP07}).
The radial velocities were determined from the lines of He\,\textsc{i} ($\lambda$4471,
$\lambda$4388, $\lambda$4144, $\lambda$4026) only. We preferred to avoid the H
Balmer lines (as did Fitzpatrick et al. \cite{FRG02}) because of moderate to strong
nebular emission polluting most systems. Four 80 \AA \ spectral ranges centered on the
four He\,\textsc{i} lines were extracted from each spectrum.
For each system, KOREL was run with a grid of values $(K_\mathrm{P}, q)$. The solution with 
the minimum sum of squared residuals as defined by Hadrava (\cite{pH04}) was retained 
as the best solution. For eccentric systems, a second run was performed letting
$K_\mathrm{P}$, $q$, $T_{0}$ and $\omega$ free to converge ($e$ is fixed by photometry).  
It is important to notice that the four spectral regions were analyzed simultaneously,
i.e in a single run of KOREL. 
Each region was weighted according to the $S/N$ of each He\,\textsc{i} line
($\mathrm{weight} \propto (S/N)^{2}$).
Beside the simultaneous retrieving of RV curves, orbital parameters and disentangled
spectra, the KOREL code is able to disentangle spectra for a given orbital solution
($K_\mathrm{P}$, $q$, $T_{0}$ and $\omega$ fixed). A final run of KOREL with this mode
was then used to disentangle the regions around the Balmer and He\,\textsc{ii} 4200 and
4542 lines. Indeed, He\,\textsc{ii} lines and a number of Si\,\textsc{iii-iv} lines are very
useful to constrain the temperature of hot components.

{\underline{\it Light curve analysis}}. 
For each system, a preliminary photometric solution was found by the application of 
the method of multiple subsets (MMS) (Wilson \& Biermann \cite{WB76}). That allowed to
provide fairly precise values of $e$ and $\omega$ that were introduced in the KOREL
analysis. Then, all three light curves and both RV curves provided by KOREL were
analyzed simultaneously using the WD code. The $I$ light curve is the most constraining
one thanks to the large number of points, while the $B$ and $V$ light curves provide accurate
out-of-eclipse $B$ and $V$ magnitudes. The mass ratio $q$ was fixed to the value found
by KOREL. The semi-major orbital axis $a$, treated as a free parameter, allows to scale
the masses and radii. In a first run, the temperature of the primary was arbitrarily
fixed to 26\,000 K. Second-order parameters as albedos and gravity darkening exponents
were fixed to 1.0. Metallicities $\left[M/H\right]$ were set at -0.5. The limb-darkening
coefficients were automatically interpolated after each fit from the Van Hamme
tables (Van Hamme \cite{VH93}). 

\begin{figure}[b]
\begin{center}
\includegraphics[height=2.3cm]{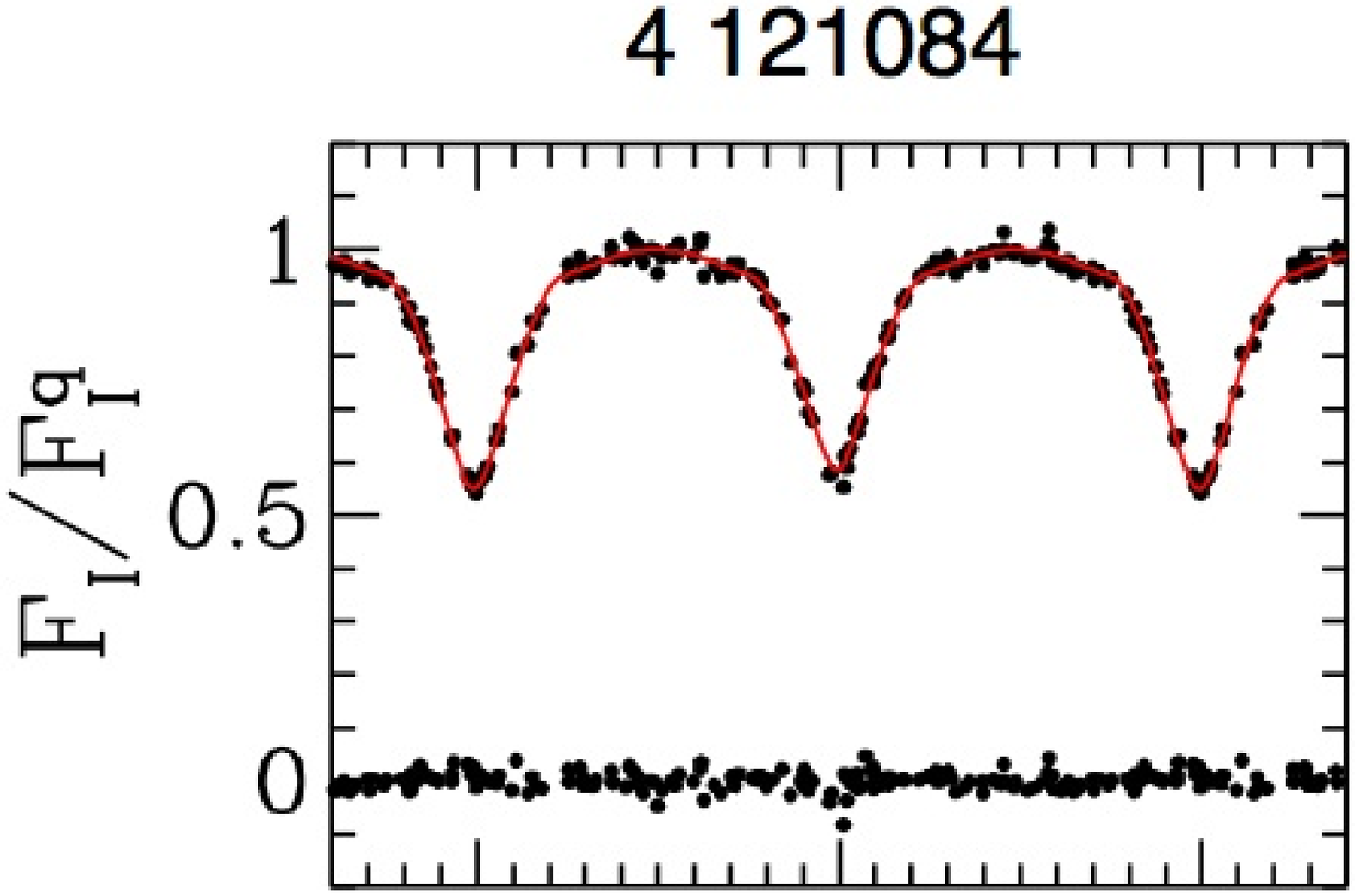}
\includegraphics[height=2.3cm]{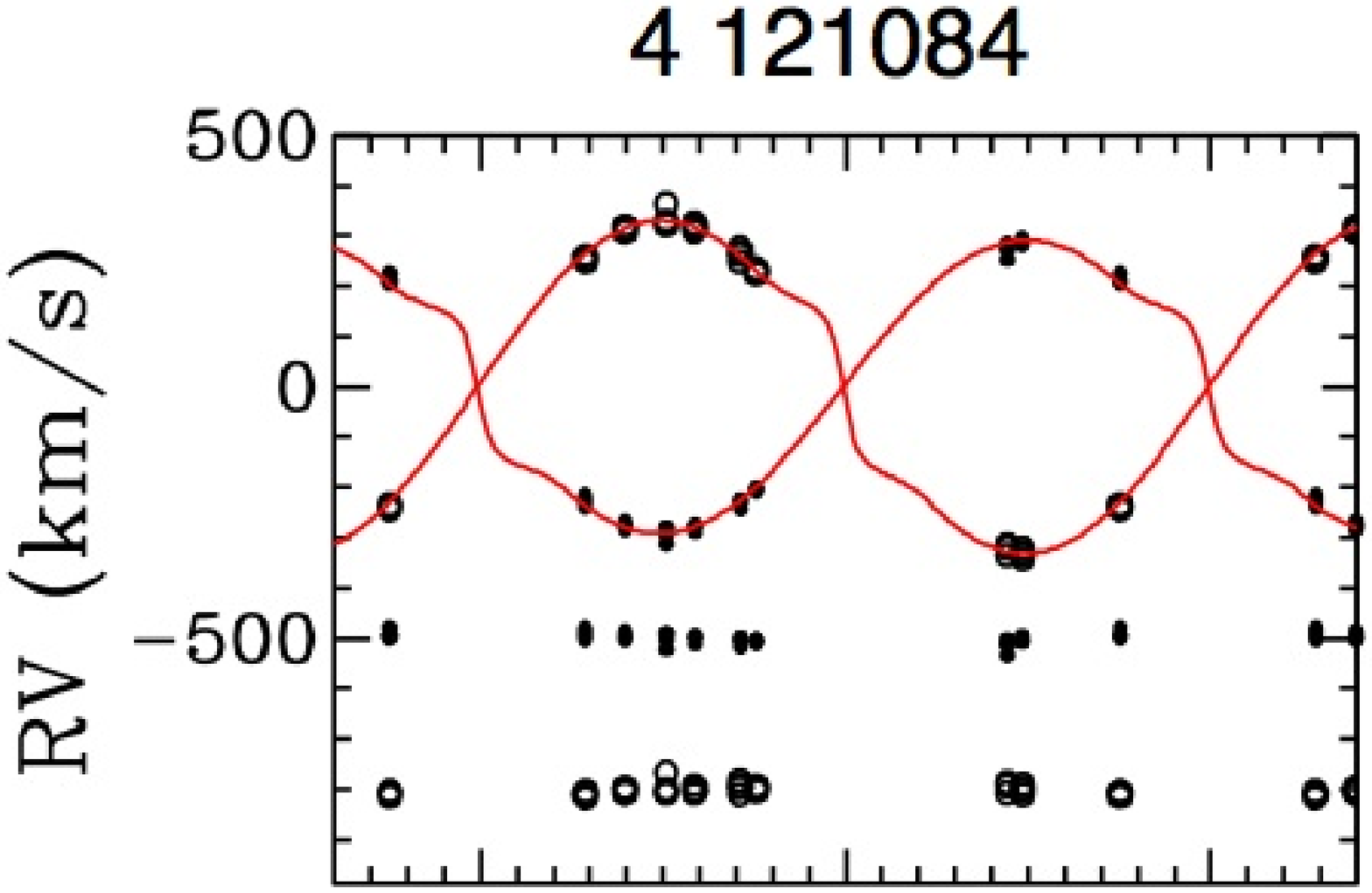}
\includegraphics[height=2.3cm]{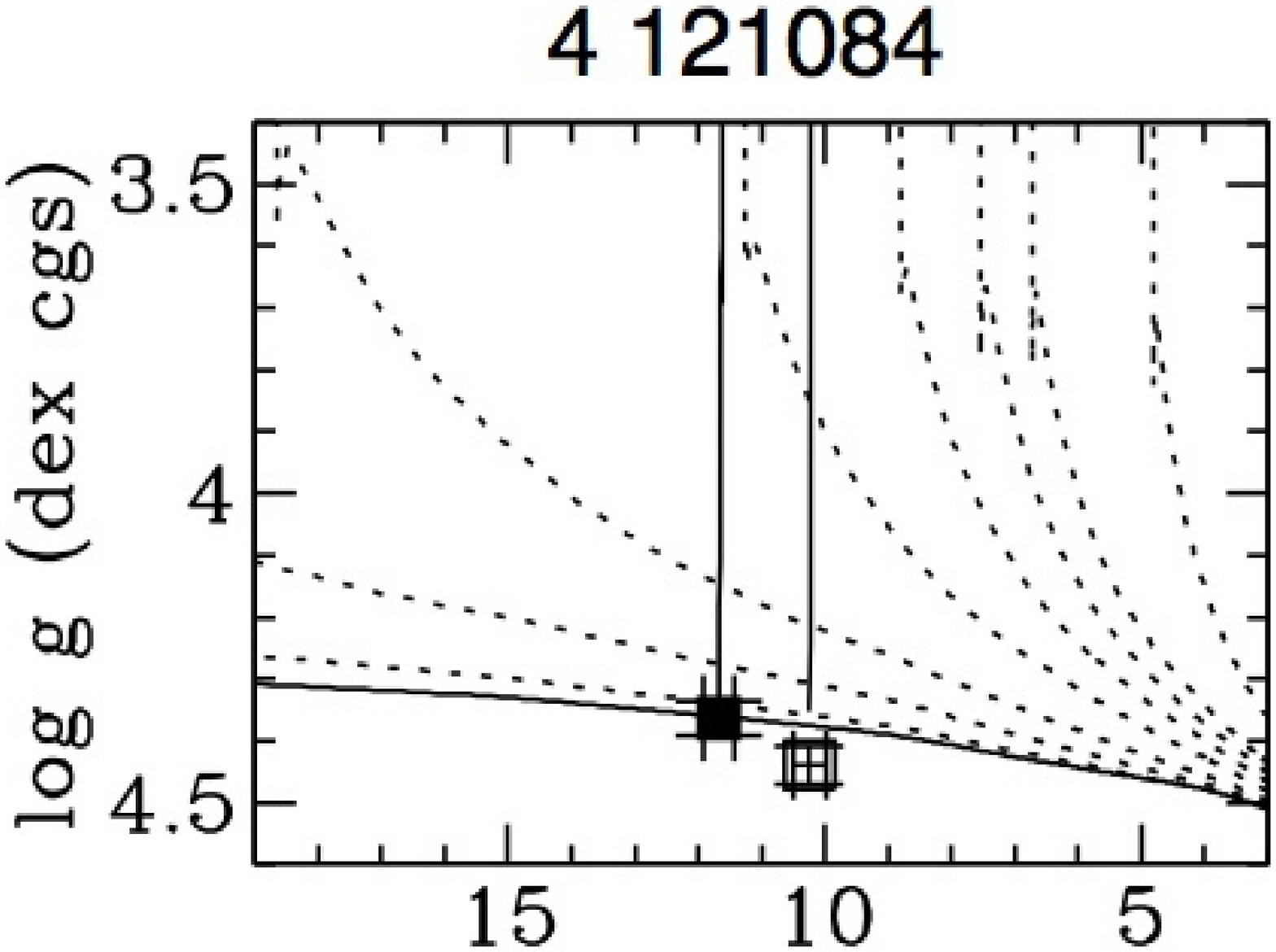}
\includegraphics[height=2.3cm]{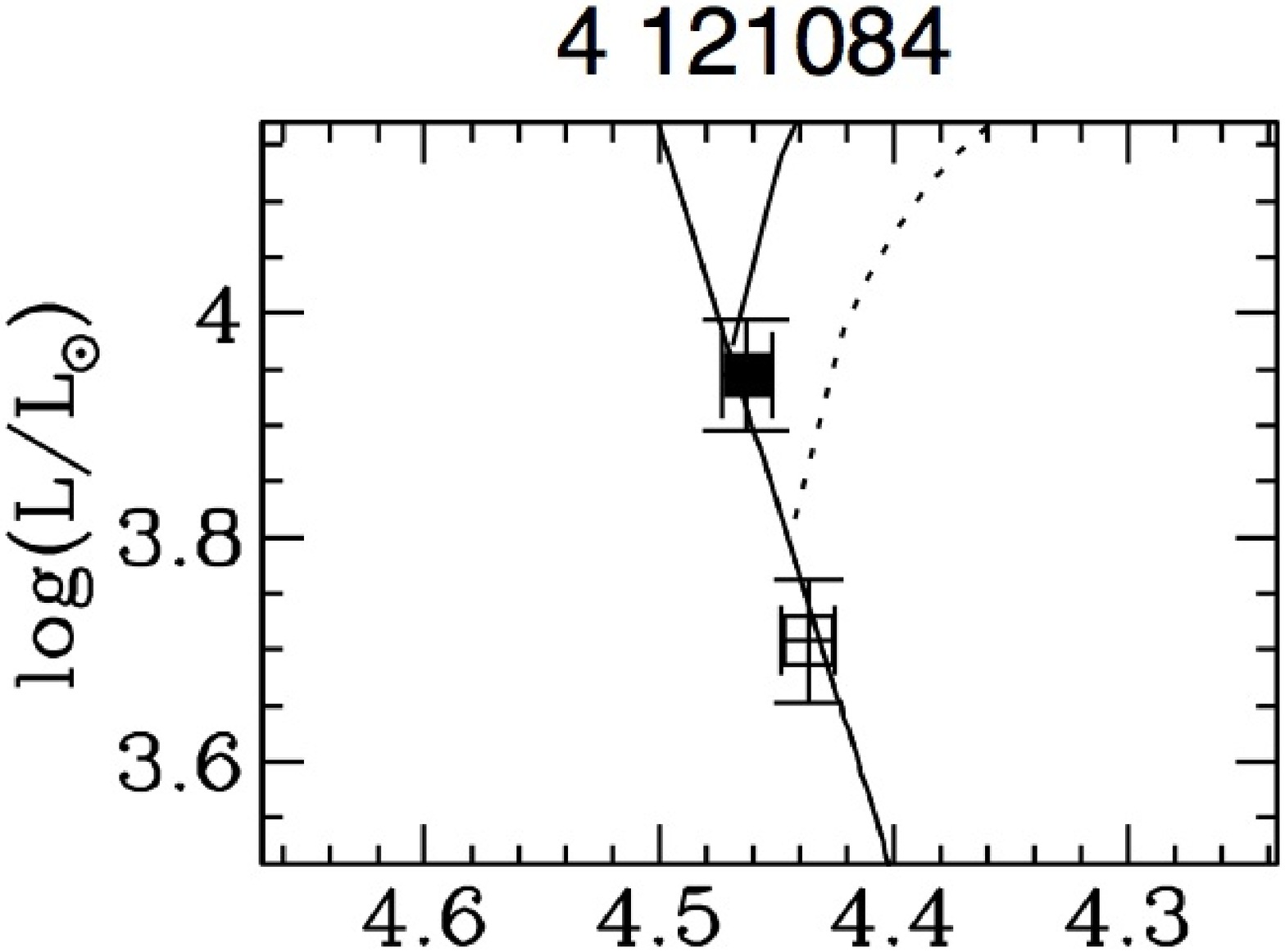}
\caption{Example of results for the binary system 4\_121084. The two left panels show
the light and velocity curves as a function of the orbital phase, the lower part of each
panel shows the difference $O-C$. The two right panels
show the surface gravity versus mass and the luminosity versus effective temperature
on a logarithmic scale (HR diagram). The isochrones shown on the 3rd panel
correspond to ages $\log t=0$ (ZAMS, continuous line), 2, 5, 10, 20, 30, 40, 50 and 100 Myr
(dotted lines).}
\label{example}%
\end{center}
\end{figure}

A fine tuning run was performed with the primary temperature found after analyzing the
observed spectra. The standard uncertainties on the whole set of parameters were
estimated in a final iteration by letting them free to converge.

\section{Results and discussion}
The four panels of Fig.\,\ref{example} illustrate the results we obtain for a typical
binary. Among our 33 systems, 23 are detached, 9 are semi-detached and 1 is a contact one.
The uncertainties range from 2 to 7\% for the masses, 2 to 20\% for the radii and 1 to 8\%
for the effective temperatures.
There is an excellent overall agreement between the empirical masses and those obtained
from interpolation of theoretical evolutionary tracks with $Z=0.004$ in the HR diagram.
The few discrepant points can be ascribed to incomplete lightcurves or to unrecognized
third light.

\begin{figure}[t]
\begin{center}
\includegraphics[height=4.0cm]{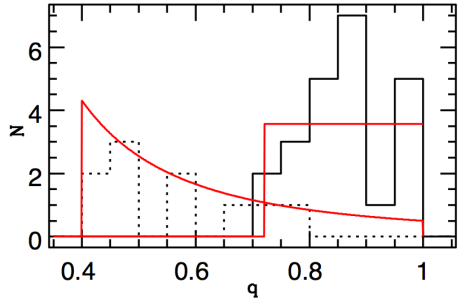}
\caption{Distribution of the 33 observed mass ratios (continuous: detached,
dashed: semi-detached/contact) with 0.05 bins. The best flat distribution (for
detached) and a Salpeter-like decreasing power law
(for semi-detached/contact) are over-plotted. Both distributions are truncated
at a cut-off value of $q = 0.72$ and $q\sim0.4$, respectively. Note that a flat
distribution would do as well for semi-detached/contact systems.}
\label{q_distr}%
\end{center}
\end{figure}

From the 21 detached systems of the HHH03/05 sample, Pinsonneault \& Stanek
(\cite{PS06}) suggest that the fraction of massive detached systems with a
mass ratio close to unity is far larger than what would be expected from a classic
Salpeter-like ($p(q) \propto q^{-2.35}$) or a flat ($p(q) = \mathrm{const}$)
$q-$distribution. While the median mass ratio is high, 0.87, there are two systems with
$q\sim 0.55$ only, which suggest that the high median value does not result from an
observational bias. On the other hand, the $q$ distribution of our sample of detached
systems does not extend below 0.7 (see Fig.\,\ref{q_distr}), which on the contrary suggests
that lower $q$ values would correspond to secondary companions too faint to be seen.
Our observed $q$ distribution is quite compatible with a flat parent distribution. For
semi-detached and contact systems, it is compatible with a flat distribution extending
from 0.4 to 0.7, or with a decreasing one, e.g. a Salpeter-like one, though without a
theoretical justification. Therefore, the twin hypothesis is clearly not substantiated
by our results, but that does not imply that the latter are compelling enough to¨
refute it. Lucy (\cite{L06}) distinguishes between the ``weak'' twin hypothesis (excess
of binaries with $q > 0.80$) and the ``strong'' hypothesis (excess of binaries with
$q > 0.95$). He shows that the strong hypothesis is verified on the basis of a sample
of 109 Galactic binary systems, and that errors larger than $\pm 0.01$ on $q$ can
smear out the signal. Therefore, the HHH03/05 sample, being much smaller and including
low precision $q$ values, is far from sufficient to confirm or deny the twin hypothesis.
Likewise, even though our sample includes undoubtedly better $q$ values, it remains
far from sufficient either.

The distance moduli are reliable for 26 systems in the $I$ band, and for 25
systems in the $V$ band.
The average distance modulus is a few hundredths of a magnitude smaller for the
$V$ than for the $I$ band. We adopt
\[ DM = 19.05 \pm 0.04 ~~~(64.5\pm 1.2~\mathrm{kpc}) \]
This value is slightly higher than that adopted by HHH05 ($18.912\pm 0.035$). The
latter authors have systems scattered over the whole SMC, which should be more
representative of the mean distance. On the other hand, our field is only
$0.45^\circ$ away from the optical centre (to the SW), so it would be difficult to
reconcile the two values by a mere geometrical effect linked with the orientation
of the SMC.

Assuming an average error of $\sim 0.1~\mathrm{mag}$ on individual distance
moduli and a gaussian intrinsic dispersion of the true moduli, a quadratic
difference yields a $2-\sigma$ depth of $0.36~\mathrm{mag}$ or 10.6~kpc.

\begin{figure}[t]
\begin{center}
\includegraphics[height=3.2cm]{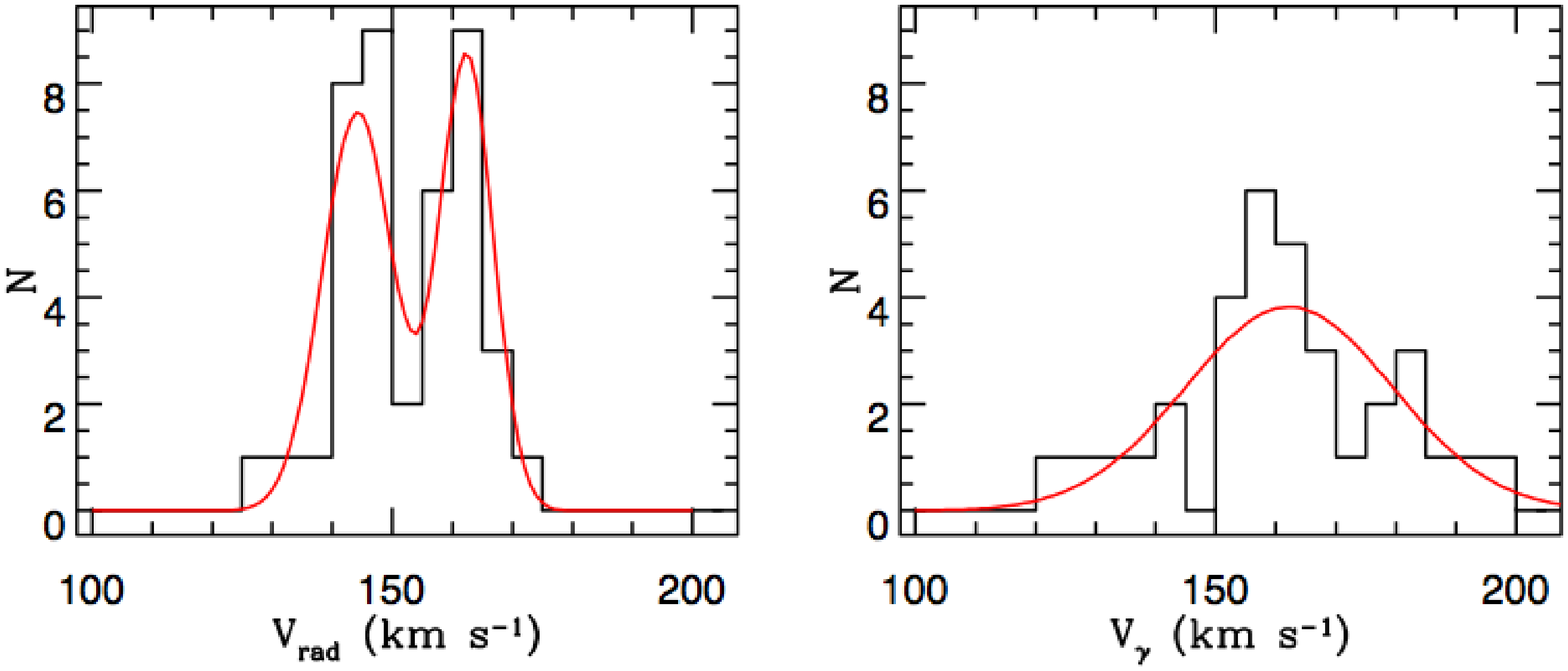}
\caption{Distribution of the radial velocities of the ionized gas (left) and of
the systemic velocities of the binary systems (right).}
\label{kinem}%
\end{center}
\end{figure}

Finally, it is interesting to compare the radial velocity distribution of the
nebular emission lines with the systemic velocities of the binary systems.
Fig.\,\ref{kinem} shows that the gas has two narrow velocity components, while
the distribution of the systemic velocities is wider and can be fit by a single
gaussian.

\end{document}